\documentstyle{mn} 
\input psfig.tex

\newif\ifAMStwofonts

\ifoldfss
  \ifCUPmtlplainloaded \else
    \NewTextAlphabet{textbfit} {cmbxti10} {}
    \NewTextAlphabet{textbfss} {cmssbx10} {}
    \NewMathAlphabet{mathbfit} {cmbxti10} {} 
    \NewMathAlphabet{mathbfss} {cmssbx10} {} 
  \fi
  \ifAMStwofonts
    \ifCUPmtlplainloaded \else
      \NewSymbolFont{upmath} {eurm10}
      \NewSymbolFont{AMSa} {msam10}
      \NewMathSymbol{\upi}     {0}{upmath}{19}
      \NewMathSymbol{\umu}     {0}{upmath}{16}
      \NewMathSymbol{\upartial}{0}{upmath}{40}
      \NewMathSymbol{\leqslant}{3}{AMSa}{36}
      \NewMathSymbol{\geqslant}{3}{AMSa}{3E}

    \fi
  \fi
\fi 

\ifnfssone
  \newmathalphabet{\mathit}
  \addtoversion{normal}{\mathit}{cmr}{m}{it}
  \addtoversion{bold}{\mathit}{cmr}{bx}{it}
  \newmathalphabet{\mathbfit} 
  \addtoversion{normal}{\mathbfit}{cmr}{bx}{it}
  \addtoversion{bold}{\mathbfit}{cmr}{bx}{it}
  \newmathalphabet{\mathbfss} 
  \addtoversion{normal}{\mathbfss}{cmss}{bx}{n}
  \addtoversion{bold}{\mathbfss}{cmss}{bx}{n}
  \ifAMStwofonts
    \ifCUPmtlplainloaded \else
      \UseAMStwoboldmath
      \makeatletter
      \new@mathgroup\upmath@group
      \define@mathgroup\mv@normal\upmath@group{eur}{m}{n}
      \define@mathgroup\mv@bold\upmath@group{eur}{b}{n}
      \edef\UPM{\hexnumber\upmath@group}
      \new@mathgroup\amsa@group
      \define@mathgroup\mv@normal\amsa@group{msa}{m}{n}
      \define@mathgroup\mv@bold\amsa@group{msa}{m}{n}
      \edef\AMSa{\hexnumber\amsa@group}
      \makeatother
      \mathchardef\upi="0\UPM19
      \mathchardef\umu="0\UPM16
      \mathchardef\upartial="0\UPM40
      \mathchardef\leqslant="3\AMSa36
      \mathchardef\geqslant="3\AMSa3E
    \fi
  \fi
\fi 

\ifnfsstwo
  \DeclareMathAlphabet{\mathbfit}{OT1}{cmr}{bx}{it}
  \SetMathAlphabet\mathbfit{bold}{OT1}{cmr}{bx}{it}
  \DeclareMathAlphabet{\mathbfss}{OT1}{cmss}{bx}{n}
  \SetMathAlphabet\mathbfss{bold}{OT1}{cmss}{bx}{n}
  \ifAMStwofonts
    \ifCUPmtlplainloaded \else
      \DeclareSymbolFont{UPM}{U}{eur}{m}{n}
      \SetSymbolFont{UPM}{bold}{U}{eur}{b}{n}
      \DeclareSymbolFont{AMSa}{U}{msa}{m}{n}
      \DeclareMathSymbol{\upi}{0}{UPM}{"19}
      \DeclareMathSymbol{\umu}{0}{UPM}{"16}
      \DeclareMathSymbol{\upartial}{0}{UPM}{"40}
      \DeclareMathSymbol{\leqslant}{3}{AMSa}{"36}
      \DeclareMathSymbol{\geqslant}{3}{AMSa}{"3E}
    \fi
  \fi
\fi 

\ifCUPmtlplainloaded \else
  \ifAMStwofonts \else 
    \def\upi{\pi}
    \def\umu{\mu}
    \def\upartial{\partial}
  \fi
\fi

\def\simlt{\lower.5ex\hbox{$\; \buildrel < \over \sim \;$}}
\def\simgt{\lower.5ex\hbox{$\; \buildrel > \over \sim \;$}}
\def\ms{M$_{\odot}$}

\def\my{M$_{\odot}$ yr$^{-1}$}

\begin{document}

\title{Chemo-spectrophotometric evolution of spiral galaxies: \\
     V. Properties of galactic discs at high redshift }

\author[S. Boissier and N. Prantzos]
       {S. Boissier and N. Prantzos \\
 Institut d'Astrophysique de Paris, 98bis, Bd. Arago, 75014 Paris}
\date{ }

\pagerange{\pageref{firstpage}--\pageref{lastpage}}
\pubyear{2000}
\maketitle

\label{firstpage}

\begin{abstract}

We explore the implications  for the high redshift universe
of ``state-of-the-art'' 
models for the chemical and spectrophotometric evolution
of spiral galaxies.  
The models are based on simple ``scaling relations'' for discs,
obtained in the framework of Cold Dark Matter models
for galaxy formation, and
were ``calibrated'' as to reproduce the properties of the Milky 
Way and of nearby discs (at redshift $z\sim0$). 
In this paper, we compare the predictions of our ``hybrid'' approach
to galaxy evolution to  observations at 
moderate and high redshift. 
We find that the models are in fairly good agreement with observations
up to $z\sim$1,
while some problems appear at higher redshift (provided there is
no selection bias in the data); these discrepancies may suggest that galaxy 
mergers (not considered in this work) played a non negligible role at $z>$1.
We also predict the existence of a ``universal''  correlation between 
abundance gradients and disc scalelengths, independent of redshift.

\end{abstract}

\begin{keywords}
Galaxies: general - evolution - spirals - photometry - abundances 
\end{keywords}

\section{Introduction}

Important progress has been made in the past few years in our understanding of 
galaxy evolution, mainly due to the number and the quality of observations of 
galaxies at intermediate redshifts (up to $z\sim$1). Observations of ground-based
telescopes, combined with results from the {Hubble} Space Telescope,
provided photometric and kinematic data on the size, luminosity and rotational
velocity of galactic discs as a function of redshift (Lilly et al. 1998, 
Schade et al. 1996, Vogt et al. 1997, Roche et al. 1998, Simmard et al. 1999). 
Although the question
of various selection effects is not settled yet, these data allow, in principle,
to probe the evolution of normal galaxies and to test  theories of galaxy
formation.

Several attempts have been made to interpret these data in the framework of
currently popular models of hierarchical growth of structure in the Universe.
Combined with simple assumptions about angular momentum conservation, such
models lead to simple scaling relations as a function of redshift and may
successfully reproduce some of the aforementionned high redshift data (Mo et al. 1998,
van den Bosch 1998, Steinmetz and Navarro 1999, Firmani and Avila-Reese 2000, Avila-Reese
and Firmani 2000).  This ``forwards'' approach 
infers the structural properties of discs from those of the corresponding 
dark matter haloes and does not consider in detail the driving force  of
galaxy evolution, namely the transformation of gas to stars; for that reason, assumptions
have to be made  about the variation of the mass to light ratio and the
colours of a disc as function of its mass and of redshift, while 
the associated chemical evolution is, in general, ignored.

On the other hand, several studies (Ferrini et al. 1994, Prantzos and Aubert 1995,
Chiappini et al. 1997, Boissier and Prantzos 1999) focused mainly on the properties
of local discs, for which a large body of observational data is available.
In those studies, the observed chemical and photometric profiles of discs are used
to constrain the  radial variation of the Star Formation Rate (SFR) and of the
infall rate. These multi-zone models naturally lead to a relation between luminosity 
and size (i.e. exponential scalelength) of discs. The fact that the Milky Way is
a typical disc galaxy helps to ``calibrate'' such models, but there is no simple
prescription as to how to extend them to other discs. 

In principle, one can utilise a ``backwards'' approach and try to  infer the
properties of high redshift discs from the histories of the models that
reproduce well the local disc population. Such an approach has been adopted
in Cayon et al. (1996), Bouwens et al. (1997), Prantzos and Silk (1998)
and Bouwens and Silk (2000).
In the latter work, a comparison is made between simplified versions of the
``forward'' and ``backwards'' approaches.

A kind of ``hybrid'' approach is adopted in Jimenez et al. (1998): they relate
the disc surface density profile to the properties of the associated dark matter
halo, while they adopt 
radially varying SFR and infall timescales such as to reproduce in detail
current profiles of the Milky Way disc.
While Jimenez et al. (1998) applied their model to  the study of Low Surface
Brightness galaxies, we developed a modified and more detailed version
of this ``hybrid'' model and applied it extensively to the study of local discs.
In a series of papers (Boissier and Prantzos 1999, 2000; Prantzos and Boissier 2000;
Boissier et al. 2001) we showed that this model can readily reproduce
a large number of local disc properties:
Tully-Fisher relations in various 
wavelength bands, colour-colour and colour-magnitude relations, gas fractions 
vs. magnitudes and colours, abundances vs. local and integrated properties,
abundance gradients,
as well as integrated spectra as a function of  a galaxy's rotational velocity. 
A crucial ingredient for the success of the model is the assumption that
the infall rate scales with mass of the galaxy, i.e. massive discs form the
bulk of their stars earlier than low mass ones.

Stimulated by the success of this ``hybrid'' model in reproducing
properties of the local disc population, we explore in this work its
implications for the high redshift Universe, comparing its predictions to
data of recent surveys.
The plan of the paper is as follows:
In Section 2, we present the models that were developped in the 
previous papers of this series in order to match the properties of
the Milky Way and of nearby spirals; we also present the adopted
probability distributions of the two main parameters of our models, namely
the disc rotational velocity $V_C$ and its spin parameter $\lambda$.
The predictions for the evolution
of spirals, deduced from those models, is shown in Section 3.
In Section 4 we compare the results of models with observations
up to $z=1$.
In Section 5, we exploire the implications at higher redshifts and we compare
our results to scalelength distributions observed up to $z=3.5$; we also predict 
that abundance gradients and scalelengths should present at all redshifts
the same correlation as the one observed locally.
The conclusions of our ``backwards'' approach for the xploration of the
high redshift Universe of disc galaxies are summarized in Section 6.

\section{Models and results for nearby discs}

In this section we briefly describe the main ingredients and the underlying asumptions
of our model for the chemical and spectro-photometric evolution of spiral
galaxies. The model has been ``calibrated'' in order to reproduce
the properties of the Milky Way disc (Sec. 2.1), then extended to other
spirals with the help of simple scaling relations that allow to reproduce
fairly well most of the observed properties of nearby spirals (Sec. 2.2).
In Sec. 2.3 we present the adopted distribution functions concerning the
two main parameters of our disc models, namely the rotational velocity
$V_C$ and the spin parameter $\lambda$.

\subsection{The Milky Way model}

The model for the Milky Way disc is presented in detail in  Boissier
and Prantzos (1999, hereafter Paper I). 
The galactic disc is simulated as an ensemble of concentric, independently
evolving rings, gradually built up by infall of primordial composition. The
chemical evolution of each zone is followed by solving the appropriate
set of integro-differential equations (Tinsley 1980), 
without the Instantaneous Recycling
Approximation. Stellar yields are from Woosley and Weaver (1995) 
for massive stars
and Renzini and Voli (1981) for intermediate mass stars. Fe producing SNIa are
included, their rate being calculated with the prescription of Matteucci and
Greggio (1986). The adopted stellar Initial Mass Function (IMF)
is a multi-slope power-law between 0.1 \ms \ and 100 \ms \ from the work of
Kroupa et al. (1993).

The spectrophotometric evolution is followed in a self-consistent way, i.e.
with the  SFR $\Psi(t)$ and metallicity $Z(t)$ of each zone determined 
by the chemical evolution,
and the same IMF. The stellar lifetimes, evolutionary tracks and spectra are
metallicity dependent; the first two are from the Geneva library 
(Schaller et al. 1992, Charbonnel et al. 1996) and the latter from 
Lejeune et al. (1997). Dust absorption is
included according to the prescriptions of  
Guiderdoni et al. (1998) and assuming a ``sandwich''
configuration for the stars and dust layers (Calzetti et al. 1994).

The star formation rate (SFR) is locally given by a
Schmidt-type law, i.e proportional to some power of the gas surface
density $\Sigma_g$ and varies with galactocentric radius $R$ as:
\begin{equation}
\label{eqsfr}
 \Psi(t,R) \ = \  \alpha \  \Sigma_g(t,R)^{1.5} \ V(R) \ R^{-1}
\end{equation}
where $V(R)$ is the circular velocity at radius $R$. This radial dependence of
the SFR is suggested by
the theory of star formation induced by density waves in spiral
galaxies (e.g. Wyse and Silk 1989). 
The efficiency $\alpha$ of the SFR  (Eq. 1) is fixed by the requirement 
that the observed local gas fraction
$\sigma_g(R_0$=8 kpc)$\sim$0.2 is reproduced at T=13.5 Gyr. 

The infall rate is assumed to be exponentially decreasing in time
with a characteristic time $\tau$. In the solar neighborhood we adopt 
$\tau$=7 Gyr in order  to reproduce the local G-dwarf metallicity distribution
(Paper I). In order to mimic the ``inside out'' formation of the disc,
$\tau$ is assumed to be  shorter in the inner zones and larger in the
outer ones.

The really ``free'' parameters
of the model are the radial dependence of the
infall timescale $\tau(R)$ and of the SFR  $\Psi(R)$.
It turns out that
the number of observables explained by the model is much
larger than the number of free parameters. In particular		
the model reproduces present day ``global'' properties 
(gas, O/H, SFR, and supernova rates), as well as	
the current disc luminosities in various wavelength bands 
and the corresponding radial profiles of gas, stars, SFR and metal abundances;
moreover, the adopted inside-out star forming scheme leads to a 
scalelength of $\sim$4 kpc in the B-band and $\sim$2.6 kpc in the K-band, 
in agreement with observations (see Paper I).

\subsection{Extension to other disc galaxies}

In order to extend the model to other disc galaxies we adopt  the 
``scaling properties'' derived by Mo, Mao and White (1998, hereafter MMW98) 
in the framework of the Cold Dark Matter (CDM) scenario for galaxy formation. 
According to this scenario, primordial density fluctuations give rise to 
haloes of non-baryonic dark
matter of mass $M$, within which baryonic gas condenses later and forms discs
of maximum circular velocity $V_C$. 
It turns out that disc profiles can be expressed in terms of only two 
parameters: rotational velocity $V_C$ 
(measuring the mass of the halo and, by assuming a 
constant halo/disc mass ratio, also the mass of the disc) and spin 
parameter $\lambda$ (measuring the specific angular momentum of the halo).
If all discs are assumed to start forming their stars at the
same time,
the profile of a given disc can  be expressed in terms of the one of our
Galaxy (the parameters of which are designated hereafter by index G):
\begin{equation}
\label{eqs1}
\frac{R_d}{R_{dG}}  \  = \  \frac{\lambda}{\lambda_G} \ \frac{V_C}{V_{CG}}
\end{equation}
and
\begin{equation}
\label{eqs2}
\frac{\Sigma_0}{\Sigma_{0G}}  \  =  \left(\frac{\lambda}{\lambda_G}\right)^{-2}
 \ \frac{V_C}{V_{CG}}
\end{equation}

Eqs. 2 and 3 allow  to describe the mass profile of a galactic disc
in terms of the one of our Galaxy and of two parameters: $V_C$ and $\lambda$.
The range of observed values for the former parameter
is 80-360 km/s  (with $V_{CG}$=220 km/s), whereas for the latter
numerical simulations give values in the 0.01-0.15 range (with $\lambda_G\sim$0.03-0.06).
Larger values of $V_C$ correspond to more massive discs and
larger values of $\lambda$ to more extended ones. 
Although it is not clear yet whether
$V_C$ and $\lambda$ are independent quantities, we treated them  as such and 
we constructed a grid of 25 models caracterised by 80 $<V_C<$ 360 km/s
and 1/3 $ < \lambda/\lambda_G <$ 3.

As discussed in Boissier and Prantzos (2000, hereafter Paper II) 
the resulting disc radii and central
surface brightness are in excellent agreement with observations, 
except for the smallest values of $\lambda$ ($\sim 1/3 \lambda_G$);
this ``unphysical'' value leads to  galaxies ressembling
to bulges or ellipticals, rather than discs. 

The two main ingredients of
the model, namely the Star Formation Rate $\Psi(R)$ and the infall time-scale
$\tau(R)$, are affected by the adopted scaling of disc properties
in the following way:

- For the SFR we adopt the prescription of Eq. 1, with the same efficiency 
$\alpha$ as in the case of the Milky Way (i.e. the SFR is not a free parameter
of the model).
In order to have an accurate evaluation of
$V(R)$ across the disc, we calculate it as the sum of the contributions 
of the disc and of the dark 
halo, the latter having a density profile of a non-singular isothermal sphere
(see Paper II).

-The infall time scale is assumed to decrease with both surface density (i.e.
the denser inner zones are formed more rapidly) and with galaxy's mass,
i.e. $\tau[M_d,\Sigma(R)]$.  In both
cases  it is the larger gravitational potential that induces a more 
rapid infall.
The  radial dependence of $\tau$ on $\Sigma(R)$ is calibrated on the Milky Way,
while the mass dependence  of $\tau$ is adjusted as to reproduce
the properties of the galactic discs. 

This model,  ``calibrated'' on the Milky Way and having as main parameter
the infall dependance on galaxy mass,  
reproduces fairly well a  very large number of properties of spiral 
galaxies at low redshift (Paper II):
Tully-Fisher relations in various 
wavelength bands, colour-colour and colour-magnitude relations, gas fractions 
vs. magnitudes and colours, abundances vs. local and integrated properties,
as well as spectra for different galactic rotational velocities. 
The main assumption of the model is
that  infall onto massive discs occured earlier than infall onto
low mass galaxies. 
More recently, in paper IV (Boissier et al., 2001)
we used a homogeneous set of observational data (a subset of the data presented in
Boselli et al., 2000) in order to test this hypothesis.
We determined the gas fraction $\sigma_G=M_{GAS}/M_{TOT}$ 
and the star formation efficiency $\epsilon$ in a large number of ``normal'' spirals
as a function of  circular velocity $V_C$.
We found that
$\epsilon$ is independent of $V_C$ while  $\sigma_G$ decreases with $V_C$. 
Taken at face value, those findings
imply that low mass discs 
had not access to their gas reservoir as early  as massive ones,
otherwise they should have already turned most of their gas into stars (because 
their current star formation efficiency is similar to  the one of massive discs,
and we assume  that this has always been the case).
This crucial finding  
justifies the assumptions we made concerning the form of the infall.

We notice that in another  recent work, 
Brinchmann and Ellis (2000) 
%
%
found  that the total stellar
mass  density in massive galaxies is constant over the redshift range $0.2<z<1$.
This implies that the stellar content of these galaxies must have formed at
higher redshifts, which  is in qualitative agreement with our
assumptions.

Finally, it is worth  noticing that the properties of abundance and colour
gradients among local spirals are readily reproduced by the 
generalization of the Milky Way model and the assumed  radial dependence
of the SFR and infall timescales
(Prantzos \& Boissier 2000, herefter Paper III).

\subsection{Distributions of the parameters $V_C$ and $\lambda$ }

\label{secDIS}

Our models presented in Sec. 2.2 simulate the evolution of individual
disc galaxies, characterised by $V_C$ and $\lambda$. In order to treat
the evolution of the disc population, we need to make some assumptions about the
probability distributions of the relevant parameters.
In this paper we assume that
the distributions $F_{\lambda}(\lambda)$ and $F_V(V_C)$ 
are independent and do not evolve in time. 
We adopt the velocity distribution suggested in 
Gonzalez et al. (2000):
\begin{equation}
F_V(V) dV = \tilde{\Psi}_* \left(\frac{V}{V_*} 
\right)^{\beta}exp\left[-\left(\frac{V}{V_*}\right)^n\right]\frac{dV}{V_*}.
\end{equation}
The parameters $\Psi_*, \ V_*, \ \beta$ and $n$ are  determined  in Gonzalez et al. (2000)
on the basis of observed Tully-Fisher relationships
and luminosity (Schechter-type) functions. 
We adopt here the set of parameters of their Table 4 (fifth row, LCRS-Courteau data) 
corresponding to the velocity interval covered by our models.
By construction, the distribution of Eq. (4) is normalised to the local luminosity
function for spirals.

The distribution of the spin parameter is given by 
\begin{equation}
F_{\lambda}(\lambda)d\lambda \ = \ \frac{1}{\sqrt{2\pi}\sigma_{\lambda}} \
exp \left[ - \frac{ln^2(\lambda/\bar{\lambda})}{2 \sigma_{\lambda}^2} 
\right]\frac{d\lambda}{\lambda}
\end{equation}
with $\bar{\lambda}$=0.05 and $\sigma_{\lambda}$=0.5, according to  numerical 
simulations (see e.g. MMW98). The distribution $F_{\lambda}(\lambda)$
is normalised to unity.

\begin{figure}
\psfig{file=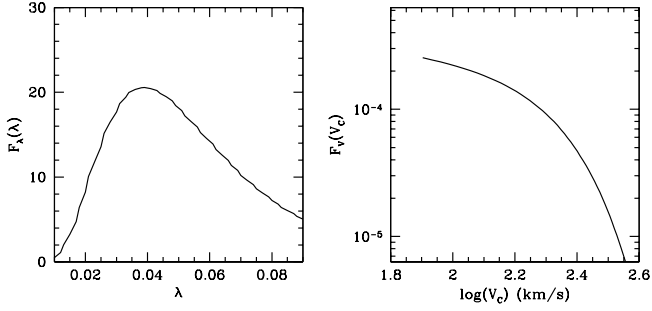,width=0.5\textwidth}
\vspace{-4.5cm}
\caption{\label{DIS} Adopted distributions of the two 
parameters caracterising our models of
disc galaxies: spin parameter $\lambda$ ({\it left}) and 
circular velocity $V_C$ ({\it right}). The $\lambda$ distribution is normalised
to $\int F(\lambda) d\lambda$=1, while  the $V_C$ distribution is normalised
to the local luminosity function in Gonzalez et al. (2000).
See text (Sec. 2.3) for details.
}
\end{figure}

Distributions $F_{\lambda}(\lambda)$ and $F_V(V_C)$ are shown in Fig. 1.
Low mass discs are more numerous than high mass ones, while 
the spin parameter distribution favours discs with
intermediate $\lambda$ values. Notice that in this work we study discs
with $V_C>$80 km/s and $\lambda<$0.1, since discs with larger $\lambda$ values
 lead to Low Surface Brightness galaxies (e.g. Jimenez et al. 1998), which
will be the subject of a forthcoming paper (Boissier and Prantzos, in preparation).

In the following, we assume that the number of discs with velocity $V_C$ and
spin parameter $\lambda$ is $dN=F_V(V_C) F_{\lambda}(\lambda) dV_C d\lambda$;
we compute then 
the distribution $\phi(Q)$ of various quantities Q (magnitude, 
radius, etc), defined by $dN= \phi(Q) dQ$.

We make the assumption  that the circular velocity and spin
parameter of a galaxy are determined at its formation  and
do not evolve in time. Since
the distributions $F_V(V)$ and $F_{\lambda}(\lambda)$ are always the same,
any evolution in  $\phi(Q)$ results from changes in intrinscic
properties of individual galaxies, according to our models.
It is clear that this assumption is an (important) 
oversimplification, since some spirals may  merge (into e.g. ellipticals),
modifying $F_V(V)$. The
history of spirals that we present concerns only fairly 
isolated dics, that have not suffered major    merger episodes.

\section{Model predictions for the evolution of spirals}

In Fig. 2 we present the results of our models concerning the evolution of the
distribution functions of various quantities: disc scalelengths $R_B$ in the
B-band (top left), central surface brightness $\mu_0(B)$  in the B-band (top right),
B-magnitude M$_B$ (middle left), colour index (U-V)$_{AB}$ (middle right),
average oxygen abundance in the gas [O/H] (bottom left) and
star formation rate $\psi$ (bottom right).
The corresponding distributions are shown at various redshifts, starting at
$z$=4.2 ({\it dashed curves}) and ending at $z$=0 ({\it solid curves}).
A cosmological model with matter density $\Omega_M$=0.3,
$\Omega_{\Lambda}$=0.7 and H$_0$=65 km/s/Mpc is adopted throughout this work.

The comparison of our results to observations of local spirals has been done in
previous papers (Papers II and III). Here we comment on the evolution of
the distribution functions $\phi(Q)$ that we obtain.
As can be seen from Fig. 2, there has been a small but steady evolution in the 
distribution of $R_B$, with all discs becoming progressively larger.
Disc scalelengths span today the range 0.8-11 kpc, compared to 0.3-6 kpc at $z\sim$4.
This is due to  the inside-out star formation scheme adopted in our models
(see Sec. 2.1 and Paper II).
The most probable value has increased only slightly between $z$=4.2 and $z$=0, 
from $\sim$1 kpc to $\sim$1.5 kpc.

\begin{figure}
\psfig{file=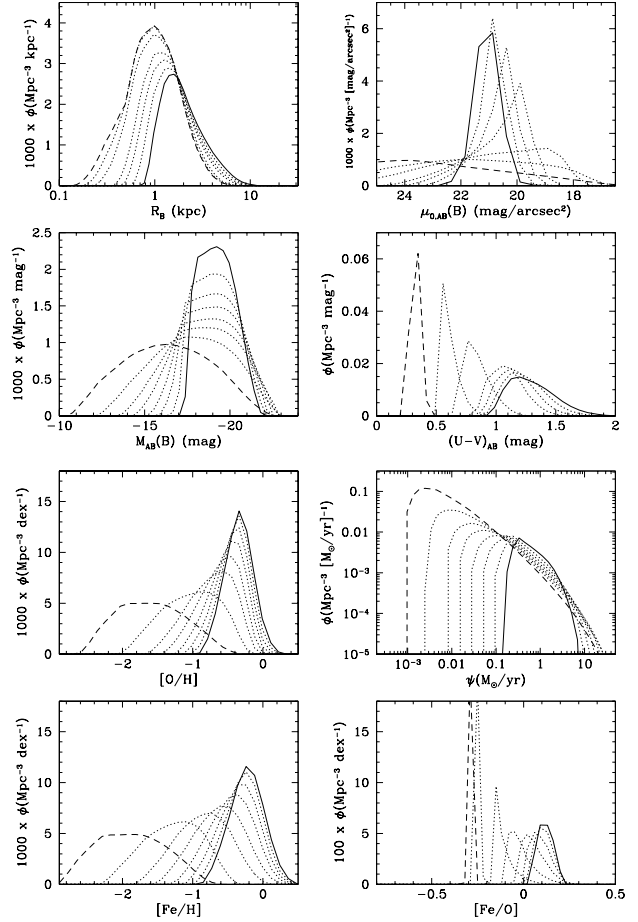,width=0.5\textwidth}
\caption{\label{DSS} Evolution of the distribution functions $\phi(Q)$
of various output parameters $Q$ of our models. Results are shown 
at different redshifts, starting with  $z$= 4.3 ({\it dashed} curves), 
then $z$=2.8, 1.7, 1.2, 0.7, 0.5, 0.2 ({\it dotted} curves) and
$z$= 0 ({\it solid} curves). 
{\it Top left:}  Distribution of disc scalelength in the B-band $R_B$;
{\it Top right:}  Distribution of central surface brightness in the B-band 
$\mu_{0,AB}(B)$;
{\it Second panel left:}  Distribution of disc restframe absolute magnitude in the 
B-band $M_{AB}(B)$;
{\it Second panel right:}  Distribution of disc colour (U-V)$_{AB}$; 
{\it Third panel  left:}  Distribution of disc average metallicity [O/H];
{\it Third panel  right:}  Distribution of star formation rate $\Psi$.
{\it Bottom left:}  Distribution of disc average metallicity [Fe/H];
{\it Bottom right:}  Distribution of disc average abundance ratio  [Fe/O].
Our models cover the range 80 $< V_C$(km/s) $<$ 360
and represent High Surface Brightness discs today (with $\mu_0(B)<$ 22.5
mag arcsec$^{-2}$), but some of them were Low Surface Brightness discs
in the past.
}
\end{figure}

The central surface brightness $\mu_0(B)$ is found to span a rather narrow
range of values today (20-22.5 mag arcsec$^{-2}$). 
Notice, however, that our models were developed to reproduce properties
of High Surface Brightness discs, while Low Surface Brightness ones
(i.e. with $\mu_0(B)>$22.5  mag arcsec$^{-2}$) could be obtained by
larger values of the spin parameter $\lambda$ (e.g. Jimenez et al. 1998).
We find that, at higher redshifts
spirals had on average higher  central surface brightness (lower $\mu_0(B)$), 
spanning a much broader range
of values. This is due to the fact that, in our scheme, massive and compact
discs evolved quite rapidly and thus developed very high central surface 
brightness early on; 
low mass and extended discs evolved slowly, starting from quite low  
values of central surface brightness 
and developing progressively brighter central regions.

The M$_B$ distribution function  was also broader at
early times, but massive and bright discs became progresssively fainter,
while low mass ones became increasingly brighter. The most probable
M$_B$ value shifted  by $\sim$-3 mag  in the past $\sim$10 Gyr.

The colour distribution function $\phi(U-V)$
shows considerable evolution in both its shape
and average value. As time goes on, galaxies become (obviously) redder, while
the increasing difference in effective ages between massive discs (formed early on)
and low mass
ones (formed much later) contributes to broaden the  $\phi(U-V)$ function.

The average metallicity is defined as the amount of oxygen in the gas divided
by the gaseous mass of the disc. It increases steadily with time
in all discs. At early times, massive discs have already developed 
relatively high O/H values, while low mass ones are essentially unevolved;
the $\phi(O/H)$ distribution is then quite broad, as in the case of $\phi(M_B)$.
At late times, massive discs show little evolution while the metallicity
of low mass ones increases rapidly, and the $\phi(O/H)$ distribution
becomes narrower.

Similar arguments hold for the distribution of the SFR $\phi(\Psi)$, which
spans a large range of values at early times, from 10$^{-3}$ \my \ to
60 \my. At late times, SFR values range from 10$^{-1}$ \my \ to 10 \my,
with an average around 1 \my.

An interesting consequence of our models is that the average Fe/O ratio
becomes slightly higher than solar at $z\sim$0, since the rate of SNIa (major
Fe producers) is such that their contribution slighly exceeds the one of O from
SNII.

\begin{figure}
\psfig{file=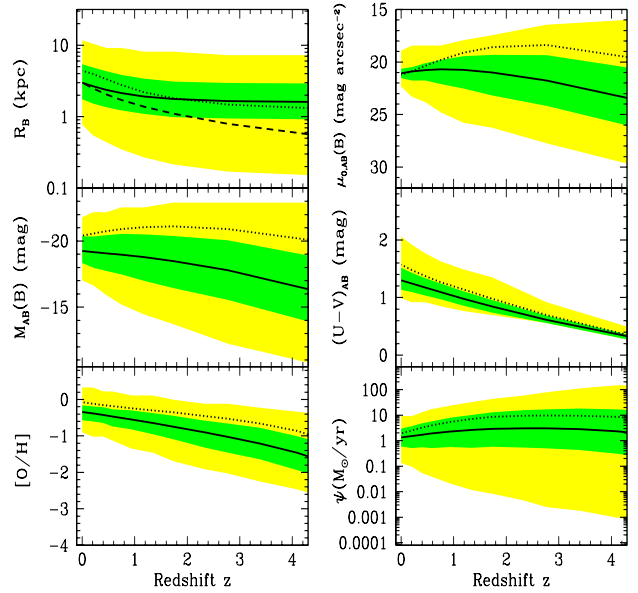,width=0.5\textwidth,height=8.5cm,angle=-90}
\caption{\label{LIL} Evolution of several quantities  computed with our models 
as a function of redshift:
scalelength $R_B$ ({\it top left}), $B$-magnitude $M_{AB}(B)$ ({\it middle left}),
average oxygen abundance ({\it bottom left}), 
central surface brightness $\mu_{0,AB}(B)$ ({\it top right}),
colour index $(U-V)_{AB}$ ({\it middle right})
and star formation rate $\Psi$ ({\it bottom right}).
In all panels the {\it solid curve} indicates the average value
of the probability distributions in Fig. 2, while the borders of the
{\it dark shaded} aerea indicate the evolution of the 
corresponding $\pm1 \sigma$ values and the {\it light shaded} aerea represents
the remaining discs.  The {\it dotted curves} indicate the corresponding
quantities for the Milky Way disc. The {\it dashed curve} in the upper left figure
indicates the predictions of hierarchical models by Mao et al. (1998), according
to which $R_B \propto (1+z)^{-1}$.
}
\end{figure}

In summary: the distributions of $\mu_0(B)$, $M_B$, O/H and $\Psi$
become narrower with time and shift to higher average values. Both
effects result from the rapid late evolution of the numerous low mass discs;
because of this late evolution, the large differences created initially by the rapid
early evolution of massive discs are reduced at late times. The distribution of
$R_B$ keeps its overall shape, but its peak value slightly increases. Finally, the
distribution of $(U-V)$  is the only one that broadens with time, because of
the increasing difference between the effective ages of massive and low mass
discs.

The results of Fig. 2 are presented in Fig. 3 in a different form. The evolution
of the average values of $R_B$, $\mu_0(B), M_B, (U-V), O/H$ and $\Psi$
(solid curves) is shown as a function of redshift $z$, 
along with the $\pm1\sigma$ values of the same quantities (inside dark shaded 
aereas). There is little increase in $<R_B>$ for $z<1$ (induced mostly
by low mass discs), while $<\mu_0>$ and $<\Psi>$ decrease slightly
at $z<2$. Other average quantities ($<M_B>, <U-V>, <O/H>$) increase
constantly, from $z\sim$4 to $z$=0. In most cases, the corresponding values
for the Milky Way (dotted curves) are larger than the average ones
(since the latter are dominated by low mass discs).

In fact, except for the case of $R_B$ and $\Psi$,  
the Milky Way values are higher than 
the average ones by more than $1\sigma$  during most of the evolution. The 
Milky Way is a large spiral and its evolution is different from the average
one not only quantitatively, but also qualitatively in some cases
(e.g. in the case of $M_B$ and $\mu_o$  where the Milky Way values 
decline for $z<2$ while the average ones continue to increase or remain
constant).

At this point, we notice that most (but not all) semi-analytical
models of galaxy evolution today use the Salpeter IMF (a power-law
with a unique slope -1.35 over the whole stellar mass range), which
is flatter than the one used here. In Paper I we discussed at length
the reasons of our choice and the current observational 
evidence against a unique slope IMF. We also mentioned briefly the main
differences resulting from its use, mainly the fact that it produces a
larger effective yield of metals and a smaller number of stars of 1-2 \ms,
which dominate the galactic luminosity at late times. Since the effective
yield of  a stellar generation depends not only on the IMF but also on the
individual stellar yields, which are still uncertain by at least a factor
of two (see Prantzos 2000 an references therein), the former effect of the
choice of the IMF cannot be tested against observations today. As for the
latter, we find that it results in differences of $\sim$30\% in luminosity
at late times, i.e. $\sim$0.15 magnitudes in the B-band; this is also 
comparable to other theoretical uncertainties entering our
calculation (neglect of late AGB phase in the adopted stellar tracks,
crude treatment of dust etc.)

\section {Evolution up to $z=1$: Models vs. Observations}

In  this section we compare our results to recent observations
concerning the evolution of  discs at moderate redshifts, for $z<1$.
A nice recent overview of the relevant observations can be found in
Hammer (1999).

\subsection{Large vs. small discs}

Lilly et al. (1998) used two-dimensional surface brightness profiles
extracted from HST images of galaxies selected from the CFRS and
LDSS redshift survey to study the evolution of several properties
of star-forming galaxies between $z=0$ and $z=1.3$.
Their observations concern the disc scale-length  $R_B$, 
magnitude $M_{AB}(B)$, central surface brightness $\mu_{AB,0}(B)$ and  colour 
index $(U-V)_{AB}$. 


The data of Lilly et al. (1998) concern  
galaxies of various inclinations, and are not corrected to the
face-on case; however, those data do not
show any trend with inclination. 
We checked  that the scatter produced by varying the 
inclination in  our models from 0 to 50 degrees is smaller than 
the scatter obtained via the variation of
the parameters $V_C$ and $\lambda$. 
We thus present our results for galaxies seen face-on.
The comparison to the data of Lilly et al. (1998) with our ``face-on''
models serves merely to check whether the ``average''
history of our models is compatible with the observational trends.

The data of Lilly et al. (1988) concern only large discs, with
scalelengths $R_B>$4 $h_{50}^{-1}$ kpc. We selected then in our models 
discs satisfying this criterion and we plotted in Fig. 4 the results
concerning the evolution of $R_B, \mu_0(B), M_B$ and $(U-A)_{AB}$
in a way similar to the one used in Fig. 3; this time, only the part
of the distribution functions of Fig. 2 concerning large discs is used
to derive average and $\pm1\sigma$ values, and results are shown only in the
$z<$1.3 range, where the measurements of Lilly et al. (1998) were made.

\begin{figure}
\psfig{file=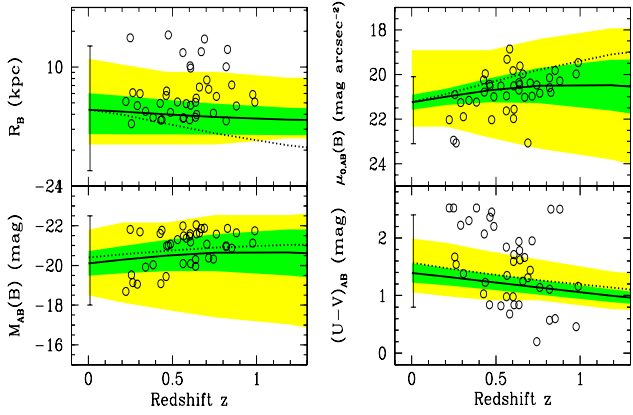,width=0.5\textwidth,height=8.5cm,angle=-90}
\vspace{-2.5cm}
\caption{\label{LIL2} Evolution of
 scalelength $R_B$, magnitude $M_{AB}(B)$,  central surface
brightness $\mu_{0,AB}(B)$ and  colour index $(U-V)_{AB}$
in the range $0<z<1.3 $.
The presentation of our results is similar to the one of Fig. \ref{LIL}, but the
average  ({\it solid curves}) and $\pm1 \sigma$ values ({\it borders of the
shaded aerea}) were
calculated taking into account only the portion of the distribution function
concerning large discs ($R_B>4 h_{50}^{-1}$ kpc).
Observations of large discs by Lilly et al. (1998) are
shown as circles. Observations of the local sample of de Jong (1996) are
inside error bars. The Milky Way evolution is shown with {\it dotted curves}.
}
\end{figure}

As can be seen in Fig.4 (left top panel) 
the data show no evolution of $R_B$ up to $z\sim$1. The
behaviour of our models is compatible with this trend, since most
large discs are already formed at $z\sim$1 and undergo very little
increase in size afterwards.
Note that our models do not reproduce the  largest 
galaxies of the Lilly et al. (1998) sample (those with $R_B>10$ kpc). 
Such discs are never obtained  in our models and we consider 
that they are not representative of ``normal'' spirals like
those of the local sample of de Jong (1996) on which we
based our modelisation in Paper II.
We notice that the Milky Way has
a more rapid late evolution than an average disc: its $R_B$ ({\it dotted curve}
in Fig. 3) decreases by a factor of $\sim$2, from $\sim$4 kpc at $z$=0 to $\sim$2
kpc at $z\sim$1.2. In this redshift range the Milky Way behaves almost 
as expected from the hierarchical scenario of Mao et al. (1998), i.e. with 
$R_B\propto (1+z)^{-1}$.

Observed B-magnitudes (left-bottom pannel) exhibit some decline
since $z\sim 1$. This is in qualitative agreement with our models
and is related to the decrease of star formation with time in massive galaxies 
(Paper II). 
The range of $M_B$ predicted by the models
is also in fair agreement with the observed one.
A comparison with Fig. 3 shows that the situation differs a lot from
the case where all discs are taken into account: in that case, the average
$M_B$ increases lately, because of the late
brightening of the numerous small discs.

There is also a slight decline of the observed central surface brightness 
of discs between $z=0$ and $z=1$ (right-top pannel), in agreement 
again with the evolution of our model $\phi(\mu_{0,AB}(B))$ for large discs.

The right-bottom pannel of Fig.  4 shows the evolution of the 
$(U-V)_{AB}$ colour 
index. The data seem to indicate a strong increase of that colour index 
with redshift (albeit with a large scatter). 
This is quite at odds with the results of our models that
predict a slow evolution in this redshift range.
However, as noticed by Lilly et al. (1998) the colour of discs in
the local sample of 
de Jong (1996) has a mean value of $(U-V)_{AB} \sim 1.5$, considerably smaller  
than their observations at moderate redshift (which suggest 
a mean value of $2$ at $z \sim 0.3$).
Notice that our models reproduce correctly colours at redshift $z=0$, since they were 
constructed as to match local observations.

The local values of de Jong (1996), combined to
the observations of Lilly et al. (1998) point towards
little colour evolution on average, as in our models.
However, the large scatter in the $(U-V)$ 
data  can not be explained by our  histories of spiral galaxies.
Brinchmann and Ellis (2000) notice that colour is a transient property. It could
be affected by small episodes of interaction, or of enhancement of star formation,
effects that are not taken into account in our models (which give an ``average'' 
history of galaxies) and could perhaps explain part of the observed scatter.

The  observations  of Lilly et al. (1998) suggest a modest amount of 
evolution of large discs in the redshift range $0<z<$1,
in agreement with the results of our models. 
This is also compatible with the conclusion of Brinchmann and Ellis 
(2000), who found that massive galaxies achieved the bulk of their
star formation before $z\sim$1.

\begin{figure}
\psfig{file=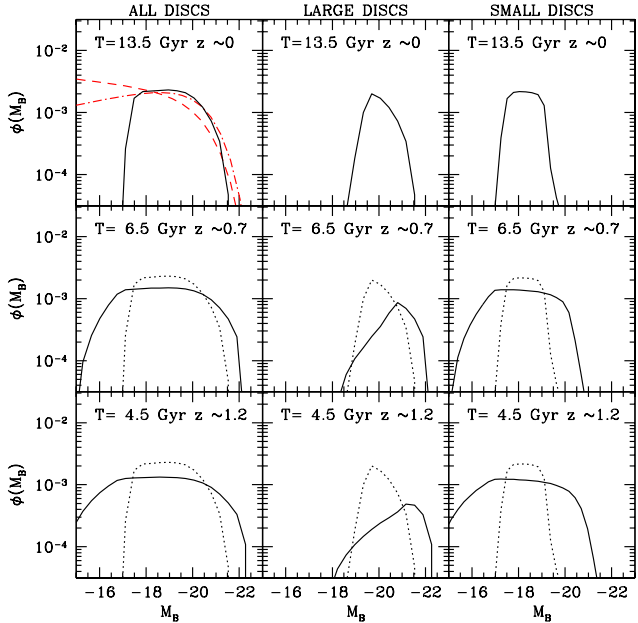,width=0.5\textwidth}
\caption{\label{fle}Evolution of the luminosity function of discs as a function of size, up
to $z \sim 1.2$.
The luminosity function for all discs is shown in the first column, while 
in the second and third columns we present  
the luminosity function of ``large'' discs
and ``small'' discs, respectively. The limit between large and small discs
 is $R_B>4 h_{50}^{-1}$ kpc, adopted  from Lilly et al. (1998).
In the top panels, the functions
are shown at redshift $z$=0 (galactic age  T=13.5 Gyr), in the middle panels at redshift
$z\sim 0.7$ (T=6.5 Gyr) and in the bottom panels at  $z\sim 1.2$ (T=4.5 Gyr). The local
($z$=0) functions of the top panels ({\it solid curves}) are repeated
as {\it dotted curves}  in the lower panels, in order to show the
evolution of $\phi(M_B)$ between low and high redshifts.
The local luminosity function for all discs of our models (top left panel)
is compared with two Schechter-type fits to observational data ({\it grey curves}), 
obtained for spiral galaxies with 
the APM survey (Loveday et al., 1992) and the SSRS2 survey (Marzke et al., 1998).
According to our models, large discs were brighter in the past, while small
discs were both brighter and fainter than today.
}
\end{figure}

Another conclusion of Lilly et al. (1998) is that the main evolutionary 
changes between $z=1$ and $z=0$ among spirals concern
small (i.e. with half-light radii ($R_{Eff} < 5 h_{50}^{-1}$ kpc)
rather than large discs. Their fig. 17 illustrates the existence of
small and bright discs at redshift $z\sim1$, 
with no local counterparts.
Our model also predicts a
stronger evolution of properties among small galaxies than among large
ones, and the existence of small and luminous spirals at high redshift.

This can be seen in Fig.  5, where we present the
luminosity function $\phi(M_B$) computed at three redshifts ($z$=0 on the top, 
$z$=0.7 in the  middle, $z$=1.2 at the bottom). The first column presents the results 
for all discs, the second column 
for ``large'' discs only (those with
$R_B>4 h_{50}^{-1}$ kpc), and the third column for  ``small'' discs
($R_B<4 h_{50}^{-1}$ kpc)
The local luminosity function (all discs, $z=0$, top left panel) 
is compared with Schechter 
functions determined for spiral galaxies of the APM survey (Loveday et al., 1992) 
and the SSRS2 survey (Marzke et al., 1998).
Taking into account that the determination of the luminosity function varies
from sample to sample, that it is affected by the choice of cosmological parameters,
and that we consider only models of face-on spiral galaxies, we think that the
agreement of the computed luminosity function with the observationally derived
ones is fairly good.

As already shown in Fig. 4, large discs were on average brighter in the past,
at least up to $z\sim$1. Fig. 5 reveals that they were also less numerous,
because some of those had  $R_B<4 h_{50}^{-1}$ kpc in  the past.
But the important feature of Fig. 5 is that small discs 
(right panels) were both fainter and brighter in the past.
A large number of small discs (comparable with their observed local
population) was brighter by $\sim$2 mag on average in the past. In our models,
those are moderately massive discs, the external parts of which are formed 
lately. At $z\sim$1 they are already bright, but not large enough. 
This population of ``small'' and bright discs at high redshift
has no counterpart today and could well constitute the
galactic population detected by Lilly et al. (1998). We further
discuss this point in the end of Sec. 4.3.

\subsection{The $R_d-V_C$ relationship}

The variation of disc size, surface brightness, magnitude etc. with
redshift can
help to test theories of galaxy formation and evolution. Mao et al. (1998)
used data at low and high redshift to deduce that 
$R_d\propto (1+z)^{-1}$, in agreement with hierarchical models
of galaxy formation.

\begin{figure}
\psfig{file=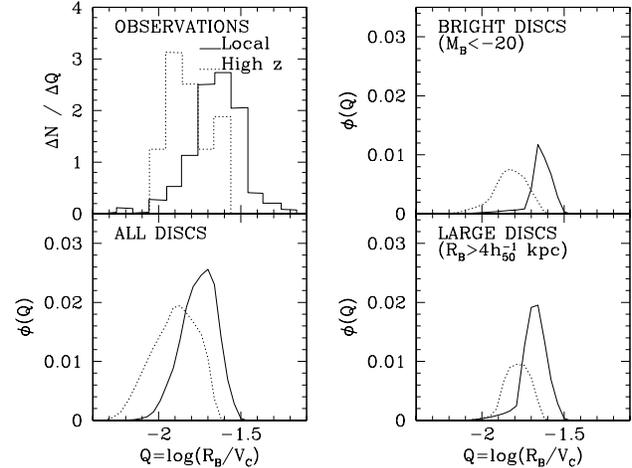,width=0.5\textwidth,angle=-90}
\caption{\label{q1}Evolution of the distribution of $Q=log(R_B/V_C)$,
where $R_B$ is in kpc and $V_C$ in km/s.
\emph{Top-left pannel}:
histograms derived by Mao et al. (1998) for a local sample (Courteau 1996, 1997,
{\it solid}), and for the high-redshift ($<z>\sim$0.7) sample of Vogt et al. 
(1996, 1997; {\it dotted}).
The comparison of the two samples suggests that discs at high redshift were smaller
for a given velocity. \emph{Bottom-left pannel}: Distribution computed with our
models at redshift $z$=0 ({\it solid curve}) and at redshift $z\sim$0.7 
({\it dotted}). We find a trend similar to the
the observationally infered one (albeit of slightly smaller magnitude), 
due to the ``inside-out'' formation of discs in our models.
\emph{Top-right pannel}: Same as bottom-left panel, for bright discs only
($M_B<-20$).
\emph{Bottom-right pannel}: Same as bottom-left panel, for large discs only
($R_B>4 h_{50}^{-1}$ kpc).
}
\end{figure}

Before presenting our results and their comparison to observations, we 
notice that the {\it size} of a disc is usually  measured by its
exponential scalelength. But a disc can also grow by simply increasing
its central surface brightness and keeping the same scalelength: it is
difficult to claim that its ``size'' remains constant in that case
(indeed, its $R_{25}$ radius increases). The use of scalelength
to measure size may  create some problems in that context.

With this caveat in mind, we proceed to a comparison between our models and
observation, concerning the quantity Q=log$(R_B/V_C)$.
The two histograms in the top-left pannel of Fig. 6 are adopted 
from Mao et al. (1998). One was computed for a local sample of discs 
(Courteau, 1996, 1997, {\it solid}),
and the other for the high-redshift ($z<1$) galaxies of Vogt et al. 
(1996, 1997, {\it dotted}). The comparison of the two histograms shows
that discs in the high redshift sample are smaller for a given velocity.
In the bottom left panel we present our results for all discs. We also find that
for a given $V_C$ discs were smaller in the past, to an extent slightly smaller
than the one inferred from observations; in fact, as can be seen in Fig. 3,
the mean value of our model $R_B$ was smaller in the past, up to $z\sim$1,
but its rate of variation was closer to $(1+z)^{-0.5}$ than to
$(1+z)^{-1}$.

These conclusions could be affected by
various bias  at high redshifts: for instance,
the study of Lilly et al. (1998) presented in Sec. 3.2 deals
only with large and bright discs, 
because fainter discs are not easily resolved.
For this reason, we present in  the right part of Fig.
6 the distribution of  Q=log$(R_B/V_C)$ 
for bright discs only ($M_B<-20$, top pannel), and for large
discs only ($R_B>4 h_{50}^{-1}$ kpc, bottom pannel). 
In both cases, the position of the maximum of $\phi(Q)$ 
is clearly shifted to lower values at high $z$, but the shift
is smaller in the case of large discs. Thus, we find that $(R_B/V_C)$ 
should be always smaller on average at high $z$, but the value of the
shift depends (albeit not very strongly) on the  sample selection criteria. 

\begin{figure}
\psfig{file=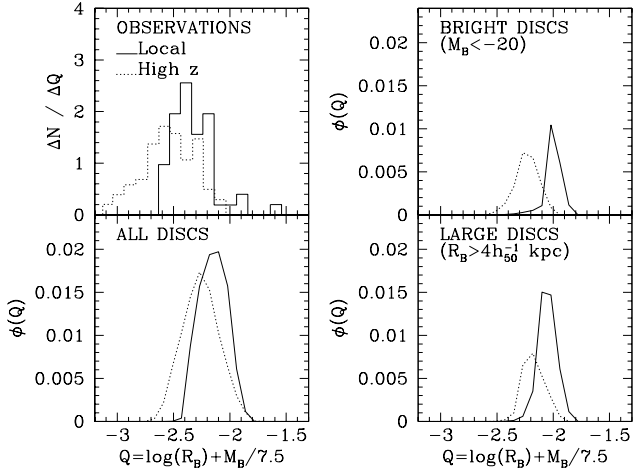,width=0.5\textwidth,angle=-90}
\caption{\label{q1}Evolution of the distribution of $Q=log(R_B)+M_B/7.5$,
where $R_B$ is in kpc.
\emph{Top-left pannel}:
histograms derived by Mao et al. (1998) for a local sample (Kent, 1985,
{\it solid}), and for the high-redshift ($<z>\sim$0.7) sample of Schade et al. 
(1996, {\it dotted}).
The comparison of the two samples suggests that discs at high redshift were smaller
for a given magnitude (or, equivalently, they were
brighter for a given size). \emph{Bottom-left pannel}: Distribution computed with our
models at redshift $z$=0 ({\it solid curve}) and at redshift $z\sim$0.7 
({\it dotted}). We find an effect compatible with
the observationally inferred one. However, the observed high $z$ distribution
is broader than ours.
\emph{Top-right pannel}: Same as bottom-left panel, for bright discs only
($M_B<-20$).
\emph{Bottom-right pannel}: Same as bottom-left panel, for large discs only
($R_B>4 h_{50}^{-1}$ kpc).
}
\end{figure}

\subsection{The $R_d-M_B$  relationship}

Mao et al. (1998) have also investigated the evolution of the  magnitude-size
relationship as a function of $z$. Fig. 7 (top-left pannel) 
shows the distribution
of the quantity Q=log$(R_B)+M_B/7.5$ that they obtained for a local sample 
(from Kent 1985) and for the high-redshift data of Schade et al. (1996).
The latter distribution is shifted with respect to the former one
towards lower values of Q.
From this comparison Mao et al. (1998) deduce that the size-magnitude relation
must have evolved since
$z \sim 1$, in the sense that discs of a given luminosity were smaller at early
times (or, equivently, discs of a given size were brighter).
The  distributions of this same quantity in our models are shown
in the left-bottom pannel, for redshifts $z$=0 and $z$=0.7, respectively. 
We also obtain a shift between the two, compatible with the one inferred 
from the observations. However, our distributions (especially the high
redshift one) are narrower than the observed ones and slightly shifted
to larger $Q$ values. The reason for this small discrepancy may be the fact
that we consider discs with $V_C>$80 km/s, i.e. slightly brighter than the
lower  limit considered in the observed samples.

The right pannels of Fig. 7 show the same quantity for bright discs (top) 
and large discs (bottom). We find again the same effect, i.e. a shift
of parameter Q towards higher values with time. Selection 
criteria do not affect then this result, namely that discs of a given
size were brighter in the past. In fact, this is another way of presenting the
result of Fig. 4, showing that large discs were slightly brighter in the
past.

\begin{figure}
\psfig{file=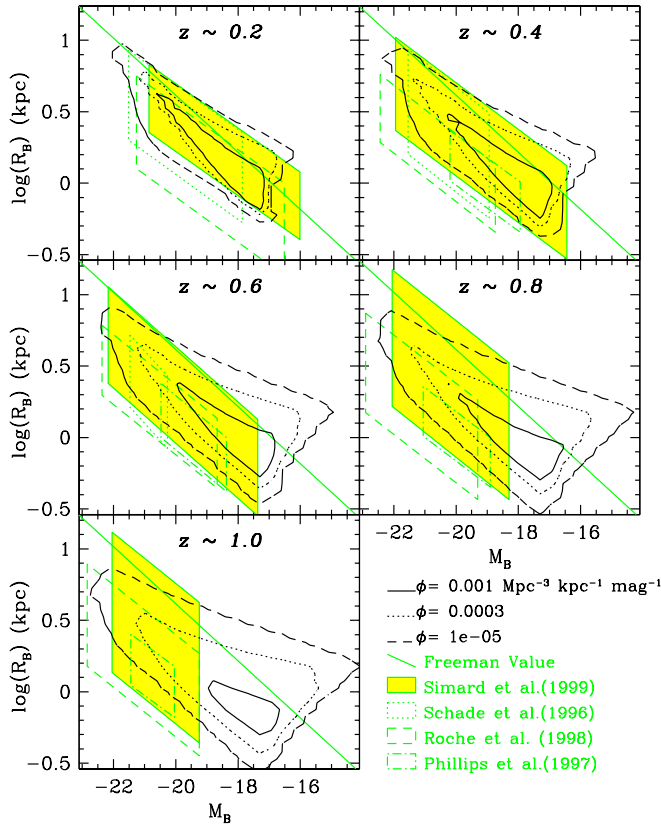,width=0.5\textwidth,height=12cm}
\caption{\label{SIZ} Disc scalelength $R_B$ vs disc rest frame absolute magnitude $M_B$
as a function of redshift $z$. 
The {\it grey boxes} indicate  various observed samples, with the one
of Simard et al. (1999) {\it shaded}; detection limits are displaced to
lower magnitudes as one moves to higher redshifts. 
Our model results, 
convolved with the $\lambda$ and $V_C$ probability functions of Fig. 1,
are shown as isodensity contours, with the largest number of discs
included inside the {\it solid} curve. At high redshifts, only the
brightest of our discs are present in the observational data.
The {\it grey diagonal line} in all panels indicates the Freeman value
of central surface brightness: $mu_{B,0}=21.65$ mag arcsec$^{-2}$ (Freeman, 1970). 
}
\end{figure}

More recently, Simard et al. (1999) analysed  190 field galaxies of the 
DEEP survey in order to determine the
magnitude-size relationship of discs in several redshift bins, up
to $z \sim 1.1$. 
Their survey is statistically complete for magnitudes $I_{814}<$23.5.
The envelope of their observations at a given redshift (taken from their paper)
is shown in Fig.
\ref{SIZ} along with a few other magnitude-size samples (Schade et al. 1996,
Roche et al. 1998, and Phillips et al. 1997), including the one
of Schade et al. (1996) that was discussed in Fig. 7.
The solid diagonal line in each panel is the locus 
of the local Freeman relation $\mu_{0B}=21.65$ mag arcsec$^{-2}$ (Freeman 1970).
To a first approximation, the data follow this relation at all redshifts,
with some  increase of the size dispersion at a given magnitude.
Also, in the highest redshift bins, the average  $\mu_{0B}$ value of the
observed samples is higher than the Freeman value. This is another way of
presenting the results of Fig. 7, namely that at high $z$ discs of a given
magnitude were smaller, on average.

Our models of $R_B$ vs $M_B$ are also plotted as a function of redshift
in the various panels of Fig. 8. We show the isodensity contours, as indicated
in the bottom-right panel of the figure.  Our results compare fairly
well to the observations of Simard et al. (1999) in the redshifts
$z\sim$0.2 and $z\sim$0.4. At higher redshifts, the dispersion of $R_B$ for a 
given  $M_B$ increases, while the average $R_B$ value decreases, both in
agreement with the observations. In fact, in the highest redshift bins,
the largest part of our model discs is fainter than the cut-off
magnitude $M_B$ of the Simard et al. (1999) survey; only our brightest
discs (the luminous ``tip of the iceberg'') are seen in the observations, 
and their behaviour is entirely compatible with the data.
The size dispersion of our discs is due to the effect of the
spin parameter $\lambda$ on their structure and evolution (through the
adopted scaling relations and the SFR and infall prescriptions, 
see Sec. 2.2). For all $\lambda$
values, massive discs attain their final size early on (before $z\sim$1) and
evolve little at late times, as discussed in Sec. 4.1. But, depending
on the $\lambda$ value, low mass discs may have very different histories:
compact ones (low $\lambda$)
evolve rapidly, i.e. before $z\sim$1, while extended ones (large $\lambda$) 
evolve more  slowly. This explains
the large dispersion in the $R_B$ values of faint discs at high redshift.

Simard et al. (1999) draw attention to the fact that the apparent
increase of mean surface brightness with $z$ in their data
may be entirely due to selection effects; if the survey selection functions
are used to correct the surface brightness distributions in the
different redshift bins, no significant evolution is found in their data up
to $z\sim$1 for discs brighter than $M_B$=-19.

This is at odds with the results of our models, which generically predict
an increase of the mean surface brightness with redshift up to  $z\sim$1.
This increase is due to the fact that most of our discs  form their central
regions long before $z\sim$1; the SFR and the corresponding surface
brightness of those regions can only decline at late times, i.e. increase
with redshift. Notice that the situation is different for $z>$1: bright discs
exhibit the same trend up to $z\sim$2 (since the evolution of their central
regions is completed earlier than that), while faint galaxies have 
mean surface brightness declining with $z$. According to our senario, they
still form their inner regions in the $z$=1-2 redshift range.

Nine galaxies at $z>$0.9 of the Simmard et al. (1999) sample
occupy a region of the $R_B-M_B$ plane rarely populated by local galaxies,
i.e. they are bright and relatively compact. In our senario, such 
high surface brightness discs do exist at $z\sim$1: they are high
$V_C$ and low $\lambda$ discs that have completed their evolution
earlier than $z\sim$1 and became progressively fainter at late times
(by $\sim$ 1 mag  in $M_B$ in the past $\sim$6-7 Gyr.

\subsection{Tully-Fisher relation}

The Tully-Fisher (TF) relation is a strong correlation exhibited 
by the whole population of disc galaxies. It relates their luminosity
$L$ or magnitude $M$ and their circular velocity $V_C$, through
$L\propto V_C^{a/2.5}$ or $M=-a[log(2V_C)-2.5]+b$.
The exact slope of the relation and its zero-point are still the subject
of considerable debate (see e.g. Giovanelli et al. 1997).

In Paper II we presented a detailed comparison of our model results
to local ($z\sim$0) TF relations obtained by different groups 
in the $I$ and $B$ bands. We noticed that in the $I$ band
(which presumably reflects better the underlying stellar population)
the TF slopes of different groups differ by more than $\sim$20\%
(i.e. from  $a$=6.80 in Mathewson et al. (1992) to 8.17 in Tully et al.
1998).

Our results are in much better agreement with the data of
Han and Mould (the corresponding TF relationship is given in Willick et al. 1996)
or of Mathewson et al. (1992), concerning
field spirals, than with those of Tully et al. (1998) or Giovanelli 
et al. (1997), which concern spirals in clusters. We notice that our
models correspond better to the former case than to the latter,
since star formation in cluster galaxies may be affected by
tidal interactions which are not taken into account in our SFR 
prescription.

\begin{figure}
\psfig{file=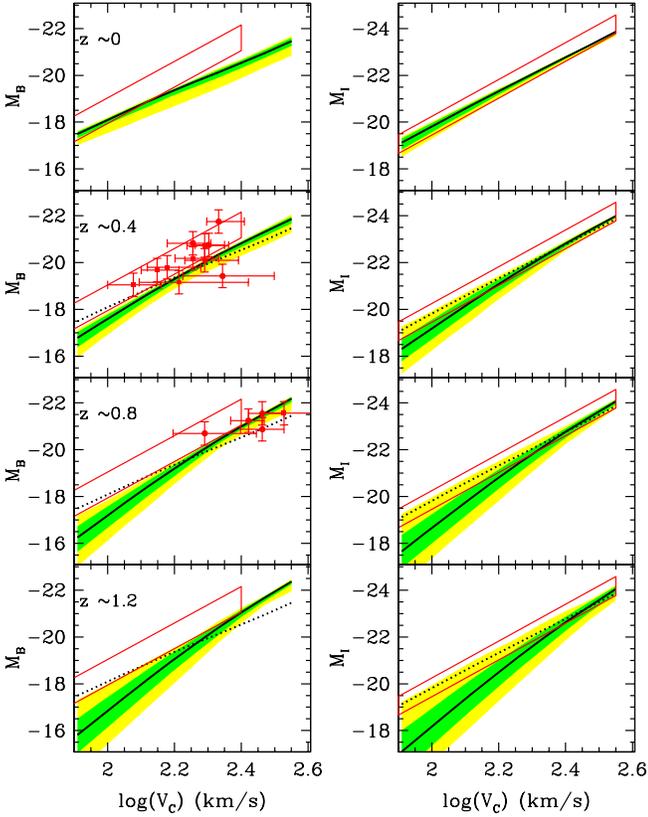,width=0.5\textwidth}
\caption{\label{TF}
Evolution of the Tully-Fisher relation as a function of redshift $z$
(from top to bottom) in the $B-$ and $I-$ bands (left and right, respectively).
The average relationship obtained with our models is shown as {\it solid curve},
among $\pm1\sigma$ values ({\it dark shaded} aerea) and all remaining
discs ({\it light shaded} aerea); in the bottom three panels, the {\it dotted
curve} is the average model TF relation of the top panel ($z$=0), allowing
to measure the evolution. {\it Boxes} in all the left panels represent 
the observed local
TF relation in the $B-$band (Tully et al. 1998), while boxes in all 
the right panels show 
the local TF relation in the $I-$band (Mathewson et al. 1996).
Data at high redshift in the left panels are from Vogt et al. (1996, 1997).
}
\end{figure}

In the $B$-band, the available local TF relation of Tully et al. (1998)
concerns again spirals in clusters. Our model $B-$band TF relation
is slightly flatter than the observed one, but in view of the
uncertainties and the caveats mentioned in the previous paragraphs,
we do not think that this a serious drawback of the models.
We note that other models (e.g. Mo et al. 1998, Heavens and Jimenez 1999, 
Bouwens and Silk 2000) apparently reproduce better
the local $B-$band TF relation. However, the mass to light ratio $M/L$
in those models is either taken as constant or it is
adjusted as a function of $V_C$ in order
to reproduce the TF relation. Our own 
$M/L$ vs. $V_C$ relation (presented in Fig. 8 of Paper II) is computed
from the fully self-consistent chemo-spectrophotometric evolution models,
which reproduce all the main properties of local discs without
any adjustment except for the initial scalings (see Sec. 2.2 and
Paper II).

Using spectroscopic observations of the Keck telescope and high-resolution
images from the Hubble deep field, Vogt et al. (1996, 1997) 
determined a Tully-Fisher (TF) relation for 
16 galaxies lying between redshifts $z$=0.15 and 1 in the rest-frame $B-$ band. 
They find no obvious change in the shape or slope of the TF relation
with respect to the local one. Assuming then the same slope,  they derive
a modest brigthning of 0.36$\pm$0.13 mag between redshifts $z\sim$0 and 1.
Their data, split in two redshift bins (0.2$<z<$0.6 and 0.6$<z<1$, respectively)
are shown in Fig. 9.

We also present in Fig. 9 the TF relationship computed with our grid
of models at four galactic ages, corresponding to $z\sim$0, 0.4, 0.8 and 1.2,
respectively (from top to bottom).
In the top panel, results are compared to the local $B-$band observations of Tully
et al. (1998). In the second and third panels, our model TF relations are compared
to the data of Vogt et al. (1996, 1997). Our results are certainly compatible with the
data, but in view of the large error bars and selection effects (only the
brightest galaxies are  detected in the the $z\sim$0.8 bin) no strong
conclusions can be drawn. It is clear, though, that our models
predict a steepening of the TF relation in the past, since massive discs
were brighter, on average, than today, while low mass discs were fainter.
This steepening with repect to the local slope is small and difficult to
detect at $z\sim$0.4 and more important in higher redshifts.
We also notice that at high redshifts the dispersion in luminosity becomes larger
for the fainter discs. In both cases,
obervations of faint galaxies ($M_B>$-18) would be required to establish
the behaviour of the TF relation  at high redshift.

\section {Evolution at redshift $z>$1.}

\subsection{Scalelength distributions}

The size-luminosity relation at redshifts $z>$1 has been recently 
investigated by Giallongo et al. (2000).
They used data from the Hubble Deep Field-North, where morphological 
information is available for all galaxies up to apparent magnitude $I\sim$26.
Using colour estimated redshifts and assuming exponential scalelengths, they
derived absolute blue magnitudes $M_B$ and disc scalengths $R_d$ as a
function of redshift. They compared these data, concerning the
redshift range 1$<z<$3.5, to those of the ESO-NTT deep field, concerning
the redshift range 0.4$<z<$1 (Fontana et al. 1999). The results of their
study appear in Fig. 10 (histograms) for ``bright'' and ``faint'' discs
(upper and lower panels, respectively).
Comparing to predictions of ``standard'' semi-analytical models,
including merging histories for dark haloes,
Giallongo et al.  (2000) found a significant excess (by a factor of $\sim$3) 
in both redshift ranges
of bright and small discs (i.e. discs with $R_d<$2 kpc in the low redshift range
and with  $R_d<$1.5 kpc in the high redshift range); they also found
a smaller excess (by $\sim$50\%) of ``faint'' discs in both redshift ranges.

\begin{figure*}
\psfig{file=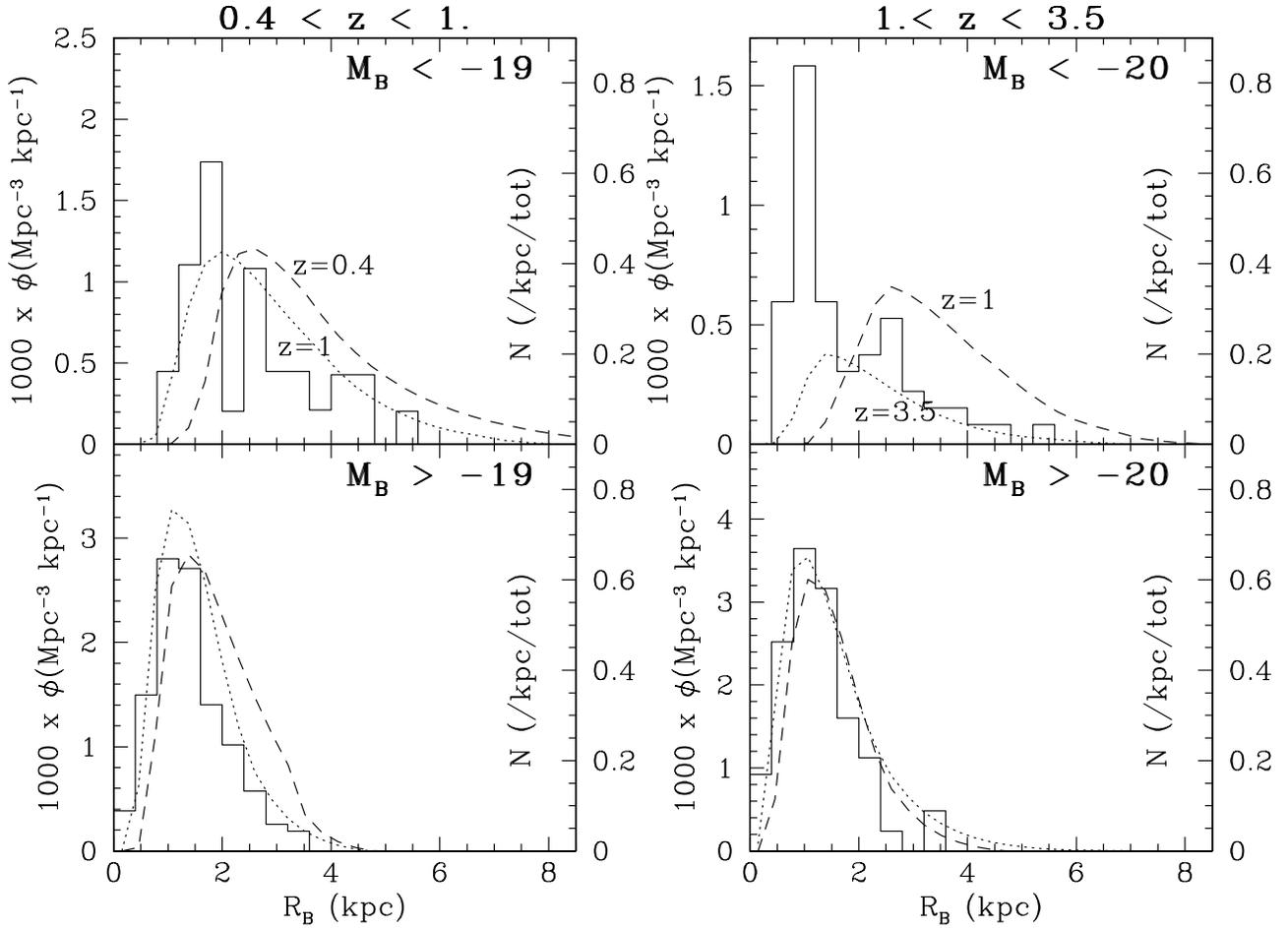,width=\textwidth,angle=-90}
\caption{\label{Q2}
Distribution functions of the scalelength in the $B-$ band $R_B$ in two
different redshift ranges: 0.4$<z<$1. ({\it left}) and 1.$<z<$3.5 ({\it right}),
respectively.
In both redshift ranges, distributions concern ``bright'' discs ({\it  upper panels}) 
and  ``faint'' discs ({\it lower panels}), respectively; notice that the magnitude limits
are different in the two cases ($M_B$=-19 on the left and $M_B$=-20 on the right).
Observed distributions ({\it histograms})
are from the  surveys of the ESO-NTT deep field at low $z$ and of the
Hubble Deep Field at high $z$ (Giallongo et al. 2000);
the corresponding scales on the vertical axis are displayed on the right of each panel.
In each panel we show two model distributions (in {\it dotted} and {\it dashed}
curves, respectively) corresponding to the two extremes of the corresponding
redshift range, i.e. at $z$=0.4 and 1 in the left panels and at $z$=1 and 3.5 in 
the right panels, respectively.
Our model distributions are not normalised to the observed
ones and the corresponding scales are displayed on the left of the vertical axes.
}
\end{figure*}

Our  results are also plotted in Fig. 10. In each panel, 
{\it dotted} and {\it dashed} curves indicate scalelength 
distributions obtained at  the two extremes of the corresponding redshift range.
When adding the model distributions of the upper and
lower panels at a given $z$ one obtains the total distribution of $R_B$ for that $z$,
shown in Fig. 2.
We notice that the observed and model distributions are  normalised 
differently (corresponding scales are indicated on the left and right axis
of each panel), but the comparison of their shapes allows to draw some 
interesting conclusions.

In the low redshift range (left panels), 
our model distributions evolve very little between $z$=1. and $z$=0.4,
with both ``bright'' and ``faint'' discs becoming larger with time.
In both cases, distributions at $z$=1 compare slightly better to the data
than distributions at $z$=0.4. The overall agreement with the data is excellent.
We find no defficiency of small discs (either bright
or faint ones), contrary to Giallongo et al. (2000). We notice that
we always use the probability distributions $F_V$ and $F_{\lambda}$
of Fig. 1, assuming no merging. This assumption turns out to be crucial
in the agreement of models with the data.
Indeed, taken at face value, the results on the left panel of Fig. 10
imply that the adopted prescriptions for the scaling relations, 
infall, SFR, $F_V$ and $F_{\lambda}$ are sufficient to make discs with the
correct distributions of scalelengths at intermediate redshifts.

In the high redshift range (right panels of Fig. 10)
we also find little evolution and excellent
agreement with observations for the ``faint'' discs ($M_B>$-20).
We notice that, by construction, our models predict 
little early evolution for discs of low mass and luminosity.
For ``bright'' discs, we find significant evolution
in the $R_B$ distribution, in the sense that this distribution
is significantly 
shifted to larger and more numerous discs when the redshift decreases
from $z$=3.5 to $z$=1: in that redshift range a large number of initially 
small and ``faint'' discs (belonging to the lower right panel of Fig. 10
at $z$=3.5) become brighter and larger at $z$=1 
(and ``migrate'' to the upper panel), 
due to the adopted inside-out formation scheme. 
However, none of the two model distributions compares
well to the observed histogram, and this will obviously be true for any
other model distribution at redshift   1$<z<$3.5: we find then
(as Giallongo et al. 2000) a serious defficiency (factor $\sim$3) of small 
and bright discs at high redshift. We notice, though, that this is the
only discrepancy of our models with respect to the
observations, while Giallongo et al.
(2000) found discrepancies in all four panels of Fig. 10.

The defficiency of our models concerns bright and small discs, i.e.
discs with high surface brightness. Selection effects may play some
role in that case, since they obviously favour the detection of
such discs, at the expense of more extended ones, having lower surface
brightness. If there is no selection bias in the data, then it is the
first hint that our models (based on the simple prescriptions of Sec. 2)
may have some difficulties in reproducing observables of the high redshift
Universe. In fact, our models are  ``calibrated'' on the Milky Way disc
and the infall timescale adjusted as to reproduce all major observables
of discs in the local Universe 
(Papers II, III and IV). The dicrepancy encountered
in Fig. 10 may imply that mergers between small discs (not taken into
account here) did play some role in that redshift range: such mergers
would affect only a small fraction of the population of the lower right panel
of Fig. 10, but would contribute significantly in populating the
defficient part of the curves in the upper rigt panel.

\subsection {Abundance gradients vs. scalelengths}

In Paper III of this series (Prantzos and Boissier 2000) we showed
that a remarquable correlation exists between the abundance gradient of 
a disc (expressed in dex/kpc) and its scalelength $R_B$: small
discs have large (in absolute value) abundance gradients and vice versa.
This is easy to understand qualitatively, since abundance gradients
are created by radially varying quantities, like the SFR $\Psi\propto$R$^{-1}$
adopted here. In a small
disc (say, $R_B$=2 kpc), such quantities vary significantly between
regions which are distant only 1 kpc from each other; a 
large abundance gradient is obtained in that case. In a large disc
(say $R_B$=6 kpc) regions separated by 1 kpc have quite similar SFR;
such discs cannot develop important abundance gradients, in dex/kpc.

\begin{figure}
\psfig{file=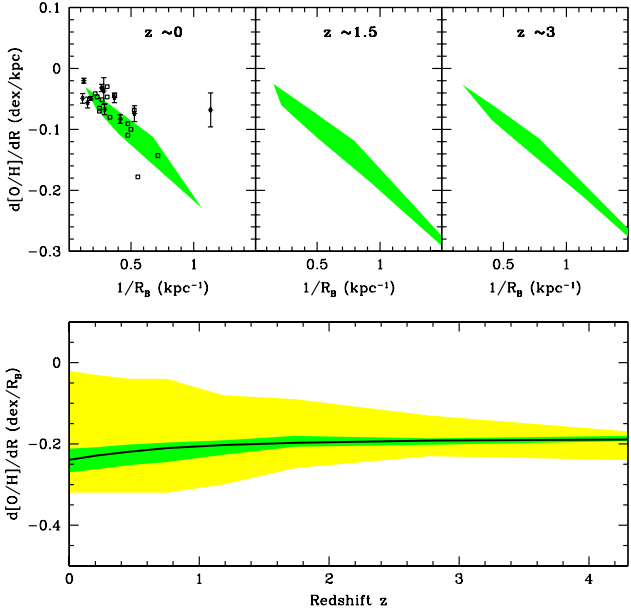,width=0.5\textwidth}
\caption{\label{grasz}
{\it Upper panels:} Relation between the abundance gradient of oxygen
dlog(O/H)/dR and the inverse of the scalelength $R_B$ as a function
of redshift. The relationship found for local discs in Paper III
(Prantzos and Boissier 2000) is supported by observations.
Data at $z\sim$0 from Garnett et al. 1997 ({\it open squares})
and van Zee et al. 1998 ({\it filled circles} with error bars); the
one dicrepant point of van Zee et al. (1998) concerns an extremely small and
low  surface brightness disc. We find a similar anticorrelation 
also  at higher redshifts; discs evolve ``homologuously''.
{\it Lower panel}: The abundance gradient of our models, expressed
in dex/$R_B$ and weighted by the probability functions $F_V$ and
$F_{\lambda}$ of Fig. 2, is plotted as a function of redshift
({\it Solid curve}: average value; {\it dark shaded aerea}: $\pm$1$\sigma$
values; {\it light shaded area:} all other discs.
}
\end{figure}

Although intuitively obvious, this property has never been properly 
emphasised before. In Paper III it was shown that
there is a unique relation between dlog(O/H)/dR and 1/$R_B$ for all
discs. This relation is found in observed nearby spirals
and is fairly well reproduced by our models of discs at $z\sim$0, as
can be seen on the upper left panel of Fig. 11.
The importance of that correlation is obvious: the abundance gradient
can be derived if  the disc scalelength is known and vice versa.

In this section we explore whether this correlation is valid at higher redshifts. 
In Fig. 11 we plot the relation of dlog(O/H)/dR vs. 1/$R_B$ for our model
discs at redshifts $z$=0 (where they compare quite favourably with observations),
$z$=1.5 and $z$=3. At all redshifts we find the relation
obtained locally. The ``homologuous evolution'' of spirals, 
invoked by Garnett et al. (1997)
for local discs, is also found in the high redshift Universe.

In the lower panel of Fig. 11 we present then the
quantity dlog(O/H)/d$R_B$ of our models (i.e. the
abundance gradient expressed in dex/$R_B$, instead of dex/kpc). We plot 
this quantity,
 weighted by the probability functions
$F_V$ and $F_{\lambda}$, as a function of redshift. It can be seen that
during most of the history of the Universe, this quantity remains $\sim$constant,
at a value of  dlog(O/H)/dR$_B\sim$-0.20 dex/$R_B$; only at very late times,
this quantity increases slightly (in absolute value), been dominated by
 the numerous small discs that   evolve considerably in the $z\sim$0.-0.5 range.
For that same reason, 
the dispersion  around this value increases at low $z$.
Both the stability of the relation of dlog(O/H)/dR vs. 1/$R_B$
and the decrease in the  dispersion of dlog(O/H)/d$R_B$ with
redshift are novel and important predictions of our model.
If true, they will allow to infer something about the chemistry of
high redshift discs from  their morphology.

At this point, we notice that Damped Luman-$\alpha$ systems (DLAs)
may be (proto-)galactic discs such as those considered
here: in a previous paper (Prantzos and Boisser 2000) we have shown that,
when observational bias are taken into account, the observations of
Zn abundance in such systems as a function of their redshift (0.5$<z<$3.5)
are nicely explained by our models. The apparent ``no chemical evolution'' picture 
suggested by the observations of those systems (e.g. Pettini et al. 1999)
is quite compatible with our understanding of the chemical evolution of discs.

\section {Summary}

In this paper we explore the implications of our disc galaxy models for the
high redshift Universe. Our models (presented in detail in Papers I, II, III and IV)
are ``calibrated'' on the Milky Way, utilise simple prescriptions
for the radial variation of the SFR and infall rate and simple scaling relations
(involving rotational velocity $V_C$ and spin parameter $\lambda$) and reproduce
all major observables concerning discs in the local Universe. A crucial ingredient
for the success of our models is the assumption that  massive discs form the 
bulk of their stars earlier than their low mass counterparts.

We assume here that the probability functions $F_V$ and $F_{\lambda}$ of the
two main parameters of our models are independent and do not evolve in time,
i.e. we assume that mergers never play a significant role in the overall evolution
of disc galaxies. We follow then the evolution of the distribution functions of
various quantities and we compare the results with available observations
concerning discs at high redshifts. Our results are summarised as follows:

1) Most large discs ($R_B>$4h$_{50}^{-1}$ kpc) have basically completed
their evolution already by $z\sim$1. Subsequently, they evolve at $\sim$constant
scalelength, while their luminosity and central surface brightness decline (Sec. 4.1).
These features of our models are in quantitative agreement with observations
of the CFH survey (Lilly et al. 1998).
Our models also predict a mild evolution of $U-V$ colour of large discs in that
redshift range, while the Lilly et al. (1998) data suggest a slightly 
stronger evolution; but their data, extrapolated to $z\sim$0, are in disagreement
with the local values derived by de Jong (1996), while our models reproduce fairly
well the observations of the local sample. Since  $U-V$ colours can be 
strongly affected by ``transient'' phenomena (e.g. short ``bursts'' of star formation),
we do not think that this discrepancy is a real problem for the models. 

2) Because of the ``inside-out'' formation scheme of discs adopted in our models,
we predict that discs were, on average, 
smaller in the past for a given rotational velocity $V_C$ or for a given  
magnitude $M_B$. Both results are in qualitative agreement with recent samples
of discs in redshifts up to $z\sim$1 (Sec. 4.2 and 4.3).
Our models reproduce fairly well the data of Simmard et al. (1999) on the
$R_B$ vs. $M_B$ relation as a function of redshift, up to $z\sim$1.
In particular, our models produce some compact and bright discs that
Simmard et al. (1999) find at $z\sim$1 and which have no counterparts 
in the local Universe (Se. 4.3).

3) We find that  the Tully-Fisher relation was steeper in the past, with
massive discs been, on average, brighter, and low mass discs been fainter
than today. Our models are in agreement with observations of the TF relation
in the $B-$ band obtained by Vogt et al. (1997) at intermediate redshifts
($z\sim$0.4 and 0.8), taking into account their error bars (Sec. 4.4). 
However, these data are insufficient to probe the
evolution of the TF relation, which requires data on fainter discs 
(down to $M_B\sim$-18 in that redshift range).

4) Our models reproduce failry well the distribution of disc scalelengths
of the ESO-NTT survey, concerning both bright and faint discs
in the 0.4-1 redshift range (Sec. 5.1). They also reproduce well
the scalelength distribution of faint discs in the 1-3.5 redshift range
obtained in the Hubble Deep Field (Giallongo et al. 2000).
On the contrary, we fail to reproduce the correpsonding distribution
of bright discs in that same redshift range: our models show a defficiency
of small ($R_B\sim$1 kpc) and bright ($M_B<$-20) discs with respect to the
Hubble Deep Field data. If this discrepancy is not explained in terms of
selection effects (favouring the detection of those high surface brightness discs)
it could be a strong indication that merging of small discs played
indeed an important role in that redshift range.

5) The anticorrelation between the abundance gradient dlog(O/H)/dR (expressed
in dex/kpc) and the scalelength $R_B$, that we found in Paper III for local
discs, is found to be valid also at higher redshifts (Sec. 5.2).
We find that the abundance gradient of discs, expressed in dex/$R_B$,
varies very little with redshift and presents a smaller dispersion at high redshifts.
These findings establish a powerful link between the optical morphology and the
chemical properties of disc galaxies, valid in both the local and the distant
Universe.

In summary, we explored the consequences of our ``backwards'' model of disc
galaxy evolution for the high redshift Universe. We found no serious
discrepancies with currently available data up to redshift $z$=1. At 
higher redshifts, we find hints that the model may require some modifications,
provided that
currently available data are not affected by selection biases.

\label{lastpage}

\end{document}